# Cuierzhuang Phenomenon：A model of rural industrialization in north China

Jinghan Tian, Jianhua Wang *

Cangzhou Normal University, Cangzhou 061001, Hebei, China

**Abstract:** Cuierzhuang Phenomenon (or Cuierzhuang Model) is a regional development phenomenon or rural revitalization model driven by ICT in the information era, characterized by the storage and transportation, processing, packaging and online sales of agricultural products, as well as online and offline coordination, long-distance and cross-regional economic cooperation, ethnic blending, equality, and mutual benefit. Unlike the Wenzhou Model, South Jiangsu Model, and Pearl River Model in the 1980s and 1990s, the Cuierzhuang Model is not only a rural revitalization brought about by the industrialization and modernization of northern rural areas with the characteristics of industrial development in the information age, but also an innovative regional economic cooperation and development model with folk nature, spontaneous formation, equality, and mutual benefit. Taking southern Xinjiang as the production base, Xinjiang jujubes from Hotan and Ruoqiang are continuously transported to Cuierzhuang, Cangzhou City, Hebei Province, where they are transferred, cleaned, dried and packaged, and finally sold all over the country. With red dates as a link, the eastern town of Cuierzhuang, which is more than 4,000 kilometers apart, connected with Xinjiang in the western region. Along the ancient Silk Road, the farthest route can reach as far as Kashgar through the southern Xinjiang route. Then, how did this long-distance and cross-regional economic cooperation channel form, what are the regional economics or economic geography principles of Cuierzhuang attracting Xinjiang jujube, and the challenges and opportunities faced by Cuierzhuang phenomenon, etc. A preliminary economic analysis has been carried out in this paper.

**Key words**: Cuierzhuang Model; Rural industrialization; Regional economic development; Rural revitalization

## 1 Introduction

In the information age, e-commerce development benefits rural areas by promoting rural industrialization, trade digitization and market networking in China. The Cuierzhuang Phenomenon is also a typical model of rural industrialization in north China, as well as a classic case of rural e-commerce driving rural revitalization in the new age. Cuierzhuang, with Cuierzhuang Town as its administrative center, is located in the west of Cang County, Cangzhou City, Hebei Province. It was previously a busting trade dock area for grain transportation on Zhujiahe River, which is connected to the Beijing-Hangzhou Grand Canal. Cuierzhuang people have become more open and flexible, and have a positive attitude towards trade as a result of canal culture. Cuierzhuang was once a small town with a long history of canal transit and trade. However, it has recently drawn our attention due to incredible industrialization and local economic development. Examining this town with a long history and splendid culture from the perspective of historical geography, using the principles of economic geography to analyze the regional development phenomenon that Cuierzhuang and Xinjiang are bound by jujube, thereby pushing the economic cooperation between the two places, and then promoting the overall development of the economy and society. Cuierzhuang Model is a new rural industrialization and modernization model with features of industrial development in the information age, compared to the Wenzhou Model, Sunan (South Jiangsu) Model, Zhujiang (Pearl River) Model, Minnan (South Fujian) Model, and Shouguang Model, etc. (Qiu and Lin, 2006; Song, 2009; Zeng et al, 2015; Su, 2020; Zhang, 2021). Theoretically and practically, analyzing the characteristics, formation, and development mechanism of this new rural industrialization and modernization model is crucial.

## 2 Cuierzhuang Overview

Cuierzhuang is located 200 kilometers south of Beijing in the North China Plain. Cuierzhuang is a township administrative unit under Cang County, Cangzhou City, Hebei Province. It is situated on the western outskirts of Cang County, close to Xian County and Hejian City. It is located between 38°15′~38°21′ and 116°27′~116°38′ with a total area of 118 km$^2$. The terrain is flat and slightly higher in the southwest. From south to north, the Zhujia Yunliang (grain transportation) River and Malanhe River drain into Beipai River. Up to June 2020, Cuierzhuang Town has jurisdiction over 48 administrative villages with a total population of 71 000, making it the largest town in Cang County. The town government is in Cuierzhuang Town's east village (Dong Cuierzhuang), 25 km east of Cangzhou City and roughly 100 km from Huanghua Port. Historically, the town was under the jurisdiction of Xian County and since 1954, it has been under the control of Cang County (Lu, 2017). Cuierzhuang township firms have sprung up in recent years, encompassing a wide range of industries including food processing, warehousing, logistics, packaging, and construction materials. There are 286 jujube processing businesses with a combined yearly output of 23000 tons of various jujube products, and more than 30 large and medium-sized cold storage facilities. Here are branches of China Mobile, China Unicom, and China Telecom. Cuierzhuang is well connected to the rest of the world thanks to its extensive land transportation network, communication network, and Internet.

## 3 Cuierzhuang Phenomenon

**3.1 Characterized by the storage and transportation, processing, packaging and online sales of agricultural products**

Cangzhou, renowned as the "hometown of Jinsi (golden-silk) jujube," has always regarded Jinsi jujube as a distinctive agricultural product. Cuierzhuang aimed to build China's largest professional jujube market as early as in 1998. It began to improve around 2010 after more than ten years of gradual development and planning, and then grew swiftly. It has now developed into an industrial agglomeration zone dominated by jujube processing, with horizontal and vertical industrial linkages including dry and fresh fruit processing, packaging, warehousing, and logistics, driving the coordinated development of multiple departments such as transportation, communication, infrastructure, catering, and tourism services. Its industrial agglomeration belt is located on both sides of the National Road G307 (also named as Qi-Yin Rd) in the south of Cuierzhuang Town, stretching about 4 kilometers from east to west.

In addition to the first planned national jujube trading market, there are some large and medium-sized cold warehouses, freight yards of freight companies, wholesale retailers of jujube and other dry and fresh fruits, hotels, logistics distribution, carton packaging, machinery supply and maintenance, etc. Furthermore, in Cuierzhuang Town's east village, west village, and surrounding villages, as well as along the National Road G105 (crossing with the National Road G307), there are many large and small processing plants, such as jujube sorting, cleaning, drying, and packaging, carton plants, color printing plants, plastic plants, automobile repair plants, and so on. Cuierzhuang Town was identified as one of the 4th national demonstration villages of *One Village One Product* (Liu and li, 2017) by the Ministry of Agriculture on July 31, 2014. Cuierzhuang Town was named the famous jujube industry town of Hebei Province in 2016 by the Department of Industry and Information of Hebei Province on April 19, 2017. Cuierzhuang was selected as Taobao town in 2020 and 2021 (No. tbt20200626, No. tbt20210747) by Ali Research Institute, and its distinctive products are jujube products and raisins.

**3.2 Online and offline collaboration driven by ICT**

Cuierzhuang people have foresight and courage, they were the first to build the country's largest red jujube trading market and played the role of so called "building a nest and attracting Phoenix," however, it is the social progress fueled by ICT (Information and Communication Technology) that really made Cuierzhuang red jujube trading prosperous. After the widespread adoption of smart phones and mobile payments in 2010, e-commerce and online businesses sprung up rapidly. The Internet and the Internet of things have brought individuals and enterprises more opportunities to connect with the outside world, show themselves and create values. Cuierzhuang is no exception.

**3.3 Long-distance and cross-regional economic cooperation**

From Cuierzhuang in Cangzhou to Kashgar in the west of Xinjiang, freight trucks always

traverse more than 4000 km across northern China. Even if the whole journey is high-speed, the fastest takes about two days, and the average takes five days. No matter how you look at it, such a long-distance cross-regional economic cooperation model violates the law of attenuation with distance in economic geography. Then, there must be another avenue for cooperation between the two, which may come through the Silk Road's continuation. After a thousand years of silence, the Silk Road which was opened by Hu-Han trade in the Han and Tang Dynasties, is resurrected now, when the Uygur and Han ethnicities are linked by jujubes. People from Xinjiang who had brought Xinjiang jujube from afar congregated in Cuierzhuang, which can hold up to one thousand people. They brought not only Xinjiang jujubes, but also Xinjiang walnuts and raisins, Ningxia medlars, Shaanxi persimmons, and even Middle Eastern coconut jujubes. Cuierzhuang has thus become a transit point for western agricultural products on their way to the east and throughout the country. Cuierzhuang has the support circumstances of technology, information, market, and transportation, while Xinjiang has the advantages as a production base. Both have mutually beneficial advantages, and win-win cooperation in jujube industry, forming a non-governmental, spontaneous, equal, and mutually beneficial economic cooperation example.

**3.4 Cultural origin of national blending**

Cuierzhuang was historically a significant trade town for grain transportation along the Zhujiahe River. Ji Yun, a prominent scholar in the Qing dynasty, was born and raised here. He was relegated to Urumqi, Xinjiang, in the 33rd year of Qianlong's reign (1768), due to his involvement in Lu Jianzeng's case of salt issues. After serving as an assistant to the military affairs there for two years, he was called back and reestablished official post, and was assigned as the head editor of the *SiKu QuanShu*. It would be a stretch to claim that Ji was the one who opened the door between Cuierzhuang and Xinjiang. However, there is another Cuierzhuanger, Zhang Zhonghan, a former deputy political commissar of Xinjiang Production and Construction Corps, the "alternative" chief of the 359th brigade and the first batch in Xinjiang military reclamation. He headed the "Bohai brigade" which established from Shandong (later the sixth division of the second army) and entered Xinjiang with General Wang Zhen in September 1949 to construct the second agricultural division of the Corps at the beginning of the People's Republic of China. From Bohai Sea to Tianshan Mountains, they reclaimed wasteland and started businesses along Kaidu river, multiplied and lived, united people of all ethnic groups, built a beautiful home, and stabilized the border of our country. Today, the beautiful new city of Shihezi whose original planning and design map was written by Zhang Zhonghan. He not only founded the Corp Art Theater and published the Corp literary magazine *Oasis*, but also founded the Corp Agricultural College (now Shihezi University) and served as the first Secretary of the CPC Party Committee. This maybe explain why Xinjiang residents eventually chose Cuierzhuang over other eastern cities.

**3.5 Regional development phenomenon or Rural revitalization model**

As early as in the 1990s, local government began to focus on the specialty Jinsi jujube, developed a base for special agricultural products, and planned to build the national largest red jujube trading market. However, after investing in infrastructure construction, it has not fulfilled its expectations in terms of jujube distribution and merchant gathering. Few merchants have come here for trading for a long time after completion. At the same time, far from Cuierzhuang, the southern Xinjiang area is reducing cotton production and expanding jujube cultivation on a large scale, and constantly introducing high-quality jujube varieties from Henan, Shandong, and Hebei. Cangzhou's technicians went to Xinjiang then. Through an initial phase from 2001 to 2010, the Xinjiang's jujube output entered a rapid growth stage after 2010 (Figure 1).

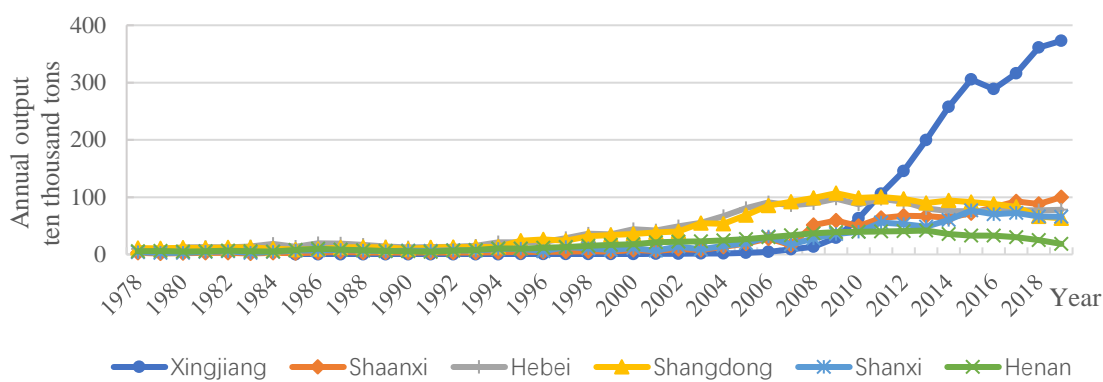

Figure 1 Annual jujube output growth in Xinjiang and traditional jujube-producing provinces from 1978 to 2019 (statistics from the National Bureau of Statistics website: https://data.stats.gov.cn/ )

In 2011, the output of dry jujube reached 1.058 million tons, surpassing traditional jujube-producing provinces such as Shandong and Hebei in the east, to become the largest jujube production base in China. In 2013, the planted area exceeded 400,000 hectares, with a full-season yield accounting for more than half of the national total (Chen et al, 2015). Xinjiang is vast, sparsely populated, and offers unique producing conditions such as plenty of light and high environmental quality. However, due to its location in the west, inconvenient transportation and underdeveloped logistics, the biggest challenge following abundant jujube harvests is sales. During this period, Cangzhou in Hebei Province, affected by rapid industrialization, has seen a dramatic fall in red jujube production since the 1990s, the peak period when its jujube output accounts for a quarter of the country. The arrival of Xinjiang jujube undoubtedly filled a gap in Cuierzhuang's jujube market. As a result, after years of adverse operations, Cuierzhuang's jujube industry and trade have finally turned around. The growth of jujube industry and trade not only drives the development of jujube storage and transportation, processing, packaging, and logistics directly, but also indirectly promotes infrastructure improvement and upgrading, as well as the prosperity and development of various industries and departments.

Therefore, this phenomenon of regional economic development is defined as follows: A regional economic development phenomenon or rural revitalization model driven by ICT in the information era, and characterized by agricultural product storage and transportation, processing, packaging, and online sales, with online and offline coordination, long-distance cross-regional economic cooperation, ethnic blending, and mutual benefit.

## 4 Economic principles of Cuierzhuang Phenomenon

### 4.1 Location selection

According to Walter Christaller's "Central Place" theory which originally published in 1933, the central place market spreads continually across the broad plain, eventually forming a hexagonal structure. The secondary center is most likely to be found at the vertex of each hexagon, which is the intersection of the three hexagons. Based on this conclusion, the emerging center is most likely to form on the outskirts of the old center market area, or the combination of the three old center market regions is the optimal location choice for the emerging center to emerge (Li, 2006). We also call it the central-marginal growth theory. Cuierzhuang is situated on the outskirts of Cang County, at the crossroads of Cang County, Xian County and Hejian City (Figure 2). However, Cuierzhuang is not the only place that follows the central-marginal growth theory. Thus, what mysterious force attracts the agglomeration of economic factors to Cuierzhuang? Then we would dig deeper into the location orientation.

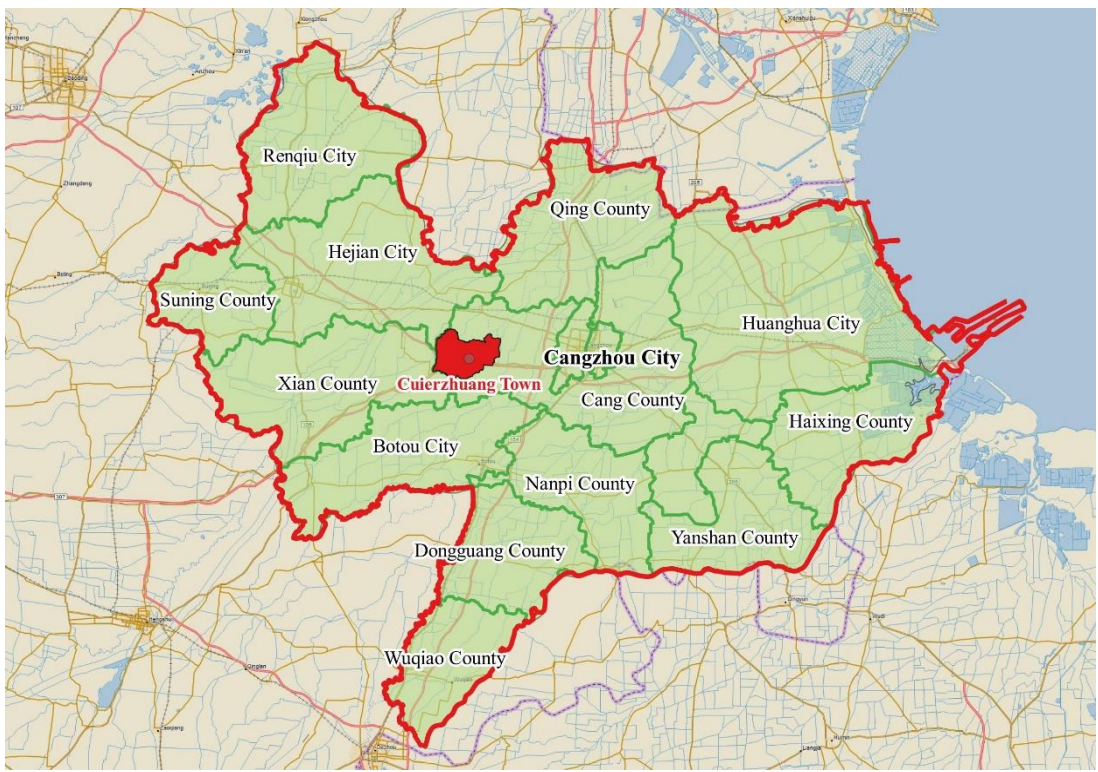

Figure 2 Location of Cuierzhuang Town, Cang County, Cangzhou City

**4.2 Location orientation**

Location orientation refers to the tendency of economic activities prefer a specific location, mainly including raw material orientation, fuel orientation, labor orientation, market orientation, transportation orientation, technology orientation, environment orientation, etc. By analyzing and comparing these many directions, it has been found that the development of Cuierzhuang is mostly driven by the humanistic environment factor. The local people are naturally sensitive to trade and have a welcoming attitude. Historically, Cuierzhuang was once an important trade town on the bank of Zhujiahe River which connected to the Grand Canal at that time. A settlement called Duoliangzhuang, located north of Cuierzhuang along the Zhujiahe River, was the official grain storage facility during the Ming and Qing Dynasties. The several Chongjiawus, commonly known as grain storage docks, are located further north and are where the government berthed grain ships (Lu, 2017) (Figure 3). Today in Cangzhou, canal transit is no longer used, and land transportation is more developed. In addition, the two locations of Cangzhou and Xingjiang are complementary in terms of technology. People in Xinjiang are skilled in trade but not good at date cultivation. Agricultural technicians from Cangzhou traveled to Xinjiang to assist in growing dates. Moreover, Cangzhou has a variety of mechanical installation and maintenance technicians which just Xinjiang lacks.

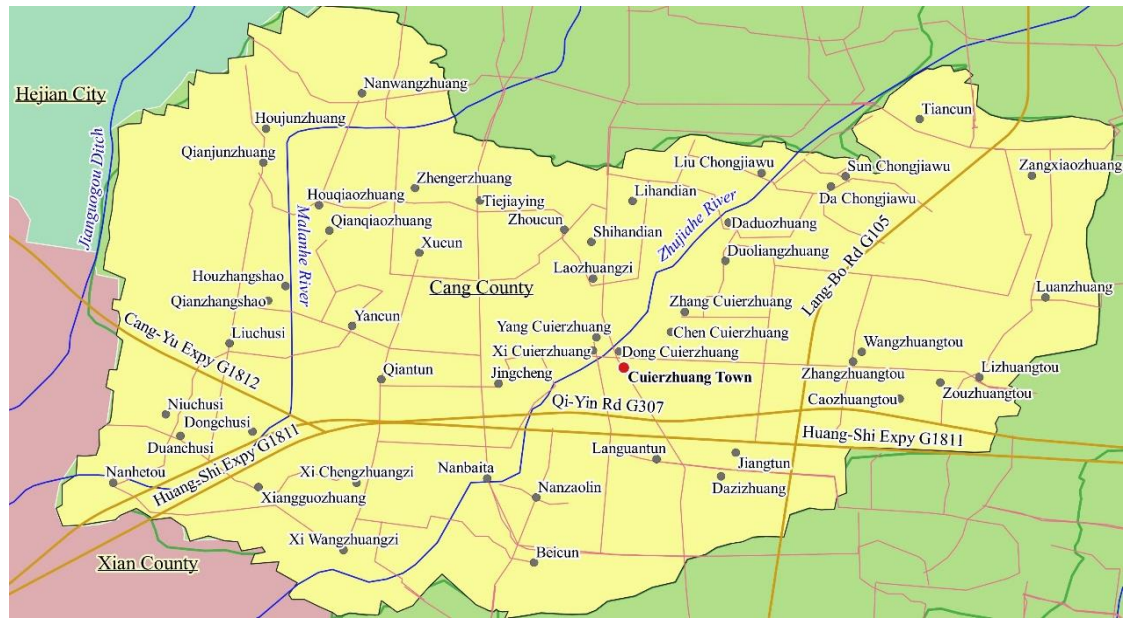

Figure 3 Transportation network of Cuierzhuang Town

**4.3 Accessibility analysis**

4.3.1 Constantly improvement of traffic conditions: Cuierzhuang is close to the entrance and exit of the expressways. The Huang-Shi Expy. G1811 (Huanghua Port to Shijiazhuang Expressway) and the Cang-Yu Expy. G1812 (Cangzhou to Yulin Expressway) cross east and west, the Jing-Tai Expy. G3 (Beijing-Taiwan Expressway) and the Jing-Hu Expy. G2 (Beijing-Shanghai Expressway) run north and south, the national highways Qi-Yin Rd G307 (Qikou-Yinchuan) and Lang-Bo Rd G105 (Langfang-Botou) meet here. The township roads are as dense as a spiderweb. Cangzhou Highspeed Railway Station is about 20 km to the east, and it takes 51 minutes by high-speed railway to travel to Beijing. The local economy has benefited much from the developed transportation network.

4.3.2 Developed information and communication network: as mentioned above, Cuierzhuang planned to build the jujube market as early as in the 1990s, but the real prosperity of the jujube market is more than ten years later. In addition to the time lag between the two places with complementary advantages in production and marketing, the limitation of ICT is also a critical factor. After 2010, with the popularity of smart-phones and the extensive rise of mobile payment and e-commerce, such a long-distance cross-regional economic cooperation has the foundation of interconnection.

4.3.3 Relevant systems and policy opportunities: Cangzhou City has implemented traffic restriction controls on yellow license trucks in recent years due to the environmental and traffic pressure, and large trucks are not permitted to enter the city. As a result, Cuierzhuang, which is not far from Cangzhou City and adjacent to the expressway entrance and exit, has the chance to become a significant truck collecting point. So, traffic congestion at Cuierzhuang, along the national highways G307 and G105, has been increasingly severe, from occasional in the beginning to common presently, practically every day and night throughout the year.

## 5 Challenges and opportunities

The original goal of Cuierzhuang's jujube market planning and construction was not to export Xinjiang jujube, but to create a distinctive agricultural product base under the Cangzhou Jinsi Jujube brand and grow it into a regional jujube production, supply, and marketing hub. This is due to the fact that in the 1990s, red jujube production in Cangzhou, Hebei Province was at its peak. However, since the strong arrival of Xinjiang jujube into Cangzhou market, Cangzhou jujube has become rare in the local market, and founding it in the domestic market has become increasingly difficult. Is it necessary to safeguard the high-quality Cangzhou Jinsi jujube in facing of this alien species invasion? And what about other high-quality jujube varieties with unique characteristics or uncommon qualities? In 2004, Cangzhou Jinsi jujube filed an application for "Protected Origin Designation". However, it appears that this is not a market-oriented, cost-effective, or successful conservation measure. On the contrary, with the slump of the Cangzhou jujube market, local villagers have begun to cut down jujube trees one after another.

Of course, the demise of Cangzhou jujube cannot be blamed on the success of Xinjiang jujube. Cangzhou jujube's exit from the market, On the other hand, is totally due to rural industrialization and urbanization. Rural industrialization has resulted in a lot of lands being abandonment and transferred to non-agricultural uses, either directly or indirectly. Rural labor is moving away from low-wage primary agricultural activities and toward higher-paying secondary and tertiary industries. Many young people congregate in factories to work or engage in individual production and operation. Apart from energy-intensive and capital-intensive heavy industries, cable accessories, pipe fittings and materials, machining, equipment maintenance, packaging materials, logistics, storage, and transportation, etc. have sprouted up everywhere in the vast rural areas of Hebei for a time, and have developed rapidly in mutual competition and cooperation, laying the groundwork for batch processing, packaging, storage, and transportation of jujube and other draught fruits. Therefore, the emergence of Cuierzhuang Phenomenon primarily benefits from the economic and social development. The countryside is being increasingly industrialized and modernized. Moreover, affected by "Path dependence", industries will continue to collect and support economic transformation and development. As Porter (1990) emphasized that the regional development is based on competitive advantages, that is, competitiveness of a less developed region is more dependent on production factors, while competitiveness of a higher developed region is from innovation (Porter, 2002; Zeibote et al, 2019). Therefore, Cuierzhuang's competitive advantage should not depend on dates or other agricultural products, but mainly based on talents and innovation in the future.

In fact, Cuierzhuang does not have ideal conditions for jujube growing. Historically, this is a low-lying salty and alkali land at the lowest tip of the Jiuhe which means "nine rivers" in Chinese. The names of adjacent villages, such as Zhangwa, Jiawa, wenwa, niuwa, and others indicate this. The "wa" refers to low-lying ground that is prone to flooding and is not suited for jujube tree planting. Only after the aggressive construction of water conservancies in the 1960s and 1970s, and after the 1980s, when the groundwater level dropped year by year and the surface water dried up, did a large stretch of dry land appear. People in Cuierzhuang noticed the change and took steps to introduce and transplant jujube trees from other villages and expand jujube planting on a large scale. Simultaneously, using local regional advantages, they planned to establish a national red jujube trading market in the 1990s, which they actively promoted and prepared the conditions for the arrival of Xinjiang jujube today. At that time, shortly after the Huang-Shi Expressway was opened, the lofty memorial archway of the traditional Cuierzhuang red jujube trading market can be seen from the freeway. Cuierzhuang is currently in the early stages of rural industrialization, which characterized by low-cost product competitiveness and a lack brand recognition. With industrialization and specialization progress, market segmentation will help to encourage industrial reorganization and adjustment. Cuierzhuang should focus more on brand creation, product quality, and reputation assurance in the future. As each rural regional development model has its applicable conditions, advantageous products, and evolutionary rules (Fang et al., 2009), therefore, how long will Cuierzhuang's jujube market and commerce last today, what higher-level industry will replace it tomorrow, and if Cuierzhuang's rural rejuvenation path is worth replicating are all unknowns. It is precisely for this reason that we should pay attention to Cuierzhuang Phenomenon and examine Cuierzhuang Model.

With informatization, networking, and digital economy empowering rural and agriculture, new rural industrialization and e-commerce have effectively promoted the circulation of agricultural products and the prosperity of the domestic market, bring opportunities for rural industrialization and rural revitalization represented by Cuierzhuang. According to a report from the Department of E-commerce and Information Technology of the Ministry of Commerce (2021), national online retail sales of rural commodities exceeded 179 trillion RMB in 2020, representing an increase of 8.9 percent year over year and 10 times the size of 2014. Agricultural items accounted for 415.89 billion RMB in online retail sales, up 26.2 percent year on year, much surpassing the 14.8 percent growth rate in national online retail sales of physical commodities during the same period (Wang and Chen, 2021). In the first half of 2021, national online retail sales of rural products totaled 954.93 billion RMB, up 21.6 percent year on year, with online retail sales of agricultural products totaling 208.82 billion yuan (Hong, 2021a), demonstrating the robust development of China's rural e-commerce as well as effectively driving the rapid growth of agricultural product sales and rural economy. Tmall Taobao accounted for 75 percent of the market share of county agricultural products in China in 2018, according to report from the Information Center of the Ministry of Agriculture and Rural Areas (2021). The transaction volume of agricultural products on Alibaba's platform was 200 billion RMB in 2019, and it exceeded 300 billion RMB in 2020, representing a 50 percent increase. As a result of the epidemic, China's e-commerce trading volume of agricultural products increased dramatically from 2020 to 2021 (Hong, 2021b). By June 2021, China's Internet users had reached 1.011 billion, while mobile Internet users had surpassed 1.007 billion, online payment users 872 million, and online shopping users 812 million (CNNIC, 2021). Therefore, Cuierzhuang should seize this historic opportunity to improve service awareness, strengthen legal concept, standardize market operation, pay attention to product quality and brand construction, establish production standards and green agricultural product certification, cultivate of professional talents, and enhance the regional competitive advantage, so as to achieve high-quality sustainable development in the face of fierce market competition.

## 6.Discussions and Conclusions

Geography emphasizes on studies of pattern and process. Pattern helps us to understand the external features of the world, and the process is conducive to the understanding of the internal mechanism (Fu, 2014; Liu, 2018). Rural industrialization and urbanization are a process that promotes the comprehensive development of regional economy and society by improving the productivity of human society. Fei (1985) was the first to put forward the "South Jiangsu Model" and pointed out that the regional development model is a distinctive economic growth path formed in a specific region under certain historical conditions (Zeng et al., 2015). By analyzing and summarizing the definitions given by scholars on the regional development models (Lin, 1987; Kong and Yang, 2011; Shang and Zeng, 2015), we propose that the regional development model for rural industrialization refers to the transformation of agricultural production mode to industrial production mode caused by changing the structure (pattern) of production factors and the operation (process) of production relations under specific temporal and spatial conditions, there by driving the overall economic and social development, with the main characteristic of a unique demonstration and reference for regional development.

Based on an analysis of Cuierzhuang Phenomenon, we summarized the definition of Cuierzhuang Model as a regional development model driven by ICT in the information age, and characterized by agricultural product storage and transportation, processing, packaging, and online sales, with online and offline coordination, long-distance cross-regional economic cooperation, ethnic blending, and mutual benefit. We also analyzed the economic principles of Cuierzhuang Phenomenon, such as location selection, location orientation, and accessibility. Finally, based on an analysis of the opportunities and challenges that Cuierzhuang face in the future, some suggestions and recommendations were offered. With the rapid development of rural e-commerce, Cuierzhuang should consider how to seize opportunities and meet challenges, shift from lower extensive production to higher intensive manufacture, and truly realize green growth, sustainable development, and rural revitalization in the future. Cuierzhuang Phenomenon makes us wonder if the Silk Road is still alive and well after a thousand years. What is the history of the connection between this grain transport river and the Grand Canal? How can the ancient canal towns preserve a sustainable and healthy development, as well as strengthen national exchanges and integration, appreciate differences, embrace diversity, and seek shared development? These are issues that deserve our attention and further investigation.